\documentclass[prl,preprintnumbers,twocolumn,superscriptaddress]{revtex4}
\usepackage{amsfonts,amsmath,amssymb,graphicx}
\begin{document}
\title{Optical spectra and exchange-correlation effects in molecular crystals}
\author{Na Sai} \affiliation{Department of Physics, The University of Texas, Austin, TX,
78712}\affiliation{Center for Computational Materials, Institute for Computational Engineering and
Sciences, The University of Texas, Austin, TX 78712} 
\author{Murilo L. Tiago}\affiliation{Materials Science and Technology Division, Oak Rdge National Laboratory, Oak Ridge,TN, 37831} 
\author{James R. Chelikowsky}\affiliation{Department of Physics, The University of
Texas, Austin, TX, 78712}\affiliation{Center for Computational Materials, Institute for
Computational Engineering and Sciences, The University of Texas, Austin, TX
78712}\affiliation{Department of Chemical Engineering, The University of Texas, Austin, TX,
78712}
\author{Fernando A. Reboredo} \affiliation{Materials Science and Technology Division, Oak Rdge National Laboratory, Oak Ridge,TN, 37831} 

\date{\today}

\begin{abstract} 
We report first-principles GW-Bethe Salpeter Equation and Quantum Monte Carlo
calculations of the optical and electronic properties of molecular and crystalline rubrene
(C$_{42}$H$_{28}$). Many-body effects dominate the optical spectrum and quasi-particle gap of
molecular crystals.  We interpret the observed yellow-green
photoluminescence in rubrene microcrystals as a result of the formation
of intermolecular, charge-transfer spin-singlet excitons.
In contrast, spin-triplet excitons are localized and intramolecular with a
predicted phosphorescence at the red end of the optical spectrum. We find that the exchange energy
plays a fundamental role in raising the energy of intramolecular spin-singlet excitons above the
intermolecular ones. Exciton binding energies are predicted to be around 0.5~eV (spin singlet) to
1~eV (spin triplet). The calculated electronic gap is 2.8 eV.  The theoretical absorption spectrum
agrees very well with recent ellipsometry data.  
\end{abstract}
\maketitle

Organic molecular crystals are promising semiconductor materials for applications in light-emitting
devices, solar cells, and electronics. Owing to the weak intermolecular interactions, transport and
photo-excitation in organic crystals deviate from those of inorganic solids with covalent or ionic
bondings~\cite{Pope,Silinsh}. Extrinsic factors such as structural imperfections and trap states
have long posed a challenge for experimental investigation of organic molecular crystals, although
recent success in fabricating organic single crystal field-effect transistors has helped to reveal
intrinsic charge transport~\cite{SCFETs}. Moreover orders of magnitude higher mobilities than in
the organic thin films or polymers have been obtained. In particular, the rubrene single crystal
has demonstrated the highest mobility~\cite{Sundar}. Currently, spectroscopy experiments including
transient absorption, luminescence, and photo-current spectroscopy have revealed dominant excitonic
effects in the optical response of the rubrene
crystal~\cite{Najafov06,Tavazzi07_JAP,Tavazzi07_PRB,Mitrofanov06}. Interpretations of the spectrum,
however, have been complicated by environmental factors such as oxidation of the rubrene
surface~\cite{Mitrofanov06,Kytka}.

On the theoretical level, a basic understanding of photo-excitations and quasi-particle gaps of a
rubrene crystal is lacking. Only limited {\it ab initio} band structure calculations within
local-density (LDA) or generalized-gradient approximation (GGA) of density-functional theory (DFT)
of rubrene have been reported~\cite{Li07,Bredas06}. The LDA and GGA are known to fail in describing
electron-hole and electron-electron interaction which are responsible for the formation of
excitons and the quasi-particle gap~\cite{Rohlfing00,Louie}. An {\it ab initio} study of the
quasi-particles and their interaction with light in a molecular crystal must include those effects. 
However, complexity has hindered the use of theories beyond DFT. We have only recently met this
challenge by a combination of advanced algorithms and parallel
computers.

\begin{figure}
\includegraphics[width=8cm]{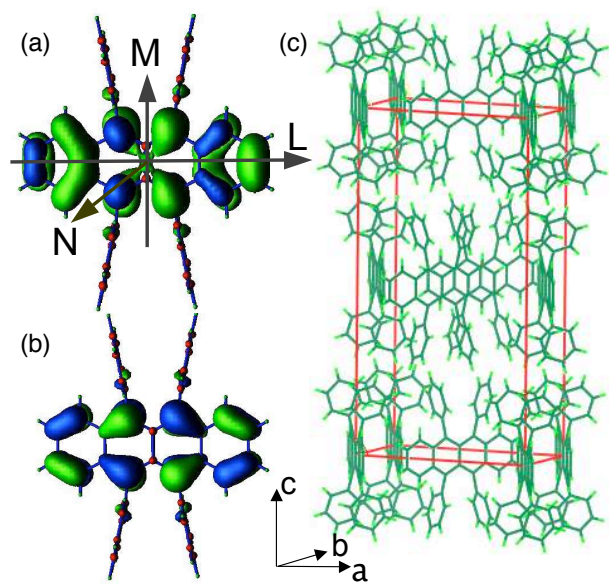} 
\caption{LUMO (a) and HOMO (b) isosurfaces in a rubrene molecule.
The tetracene backbone is on the plane of the figure. Green and blue 
colors denote positive and negative values. (c): Schematic view of the rubrene crystal.}
\label{structure}
\end{figure}
In this letter, we demonstrate the possibility of {\it ab initio} calculation of excitations
including correlation and excitonic effects in technologically relevant molecules such as rubrene.
We find that the exchange interaction plays a fundamental role in determining the properties of
excitations in this material. 

Our approach consists of four steps: i) The ground state
structures and the electronic wavefunctions of the system are described within
pseudopotential DFT-GGA; ii) we calculate quasi-particle energies within the GW approximation,
which includes many-body effects in the electron self-energy~\cite{Louie}; iii) the optical
excitations are obtained by solving the Bethe-Salpeter equation (BSE) for electrons and
holes~\cite{Rohlfing00} and finally, iv) we assess the validity of the GW-BSE methodology by
comparing its prediction of ionization potential and the first spin-triplet excitation with
diffusion Quantum Monte Carlo (QMC) calculations~\cite{QMC}. 
 
We start by exploring the electronic properties of individual rubrene molecules. The isolated
molecule has $C_{2h}$ symmetry. It consists of 4 phenyl groups attached to a tetracene backbone. 
Fig.~\ref{structure} shows isosurfaces of the highest occupied molecular orbital (HOMO) and the
lowest unoccupied molecular orbital (LUMO) of the molecule. $L$, $M$, and $N$ denote the long,
short and normal axes of the tetracene backbone, respectively. HOMO $-$ LUMO electronic dipole
transitions along directions $L$ and $N$ are dipole-forbidden. The lowest dipole-allowed transition
is polarized along the $M$ axis.

Within the GW approximation for finite systems~\cite{Tiago06}, the ionization energy $E_I$ and the
electron affinity $E_{A}$ for the isolated rubrene molecule are respectively 6.30~eV and 1.88~eV. 
Independent diffusion QMC calculations~\cite{QMC} give $E_I\sim6.22\pm$0.14 and
$E_{A}\sim0.96\pm$0.14~eV. The calculated ionization energies compare well with 6.4~eV from the
ultraviolet photo-electron spectroscopy~\cite{Sato81}, which validates the method for this system.
The overestimation of electron affinity within the GW approximation compared with QMC has been also
observed in fullerenes in gas phase~\cite{fullerenes}. The accuracy of energy gaps calculated
within the GW approximation seems to be much better in organic solids \cite{Tiago03,Shirley93}.
\begin{figure}
\includegraphics[width=8cm]{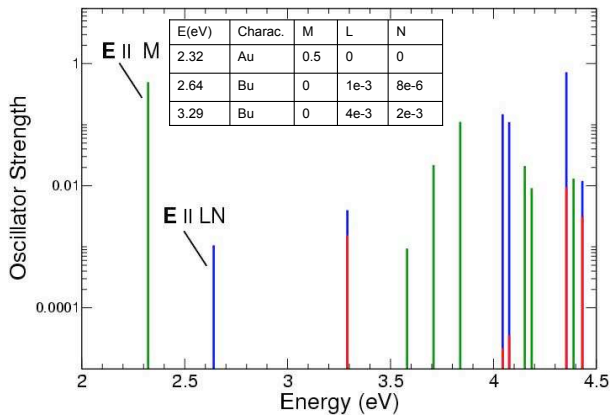}
\caption{Calculated oscillator strength of the first few optical transitions in gas-phase rubrene.
The inset shows the excitation energies, group symmetry characters, and oscillator strengths of the
three lowest excitations.}
\label{osc_mol}
\end{figure}

Absorption lines obtained within the BSE are shown in Fig.~\ref{osc_mol}.  The first absorption
line is at 2.32~eV, polarized along the $M$ axis. This is in good agreement with the experimental
absorption peak at 2.35~eV of the rubrene molecule in
solution~\cite{Tavazzi07_PRB}. The next
absorption line is at 2.6~eV, polarized in the $LN$ plane. Both excitation energies obtained from
our calculations are $\sim$ 0.5~eV lower than those calculated by semi-empirical methods
~\cite{Tavazzi07_PRB}. At higher energy, there is an intense absorption line at 4.35~eV, consistent
with experimental data (4.1~eV)~\cite{Otomo02} and semi-empirical calculations
(4.48~eV)~\cite{Tavazzi07_PRB}. 

To characterize better the electronic states contributing to the excitons, we project the BSE
eigenvectors onto each energy band. The first excitation is dominated by the transition between the
HOMO and LUMO with about $89\%$ weight. The remaining $11\%$ comes from energy levels further away
from the HOMO-LUMO gap. The second BSE excitation is strongly mixed, with components involving the
7 highest occupied energy levels and $\sim 11$ unoccupied levels. The HOMO-(LUMO+1) transition
itself contributes only $30\%$ to this excitation.  Besides the spin-singlet states in the
absorption bands, spin-triplet states which are dipole-forbidden can also be obtained. The
excitation energy is lower for the triplet, as a result of absent exchange repulsion among
electrons and holes. From GW/BSE, we have obtained an excitation energy of 0.8~eV for the first
triplet state, which should be compared with 1.45$\pm$0.13~eV predicted from our QMC calculations.
The discrepancy of 0.65~eV originates from the difference in the electron affinity calculated by
the two different methods.

Rubrene crystallizes in an orthorhombic structure with lattice parameters $a= 14.4$ \AA, $b = 7.2$
\AA, and $c= 26.8$ \AA~\cite{Jurchescu06}. Each unit cell contains two staggered $ab$ planes
separated by $c/2$. Each plane contains two translationally inequivalent rubrene molecules arranged
in a herringbone structure (see Fig.~\ref{structure}c). Across planes, the molecules are bonded by
weak forces. As a result, the electronic properties of the material are not sensitive to the exact
stacking of layers. Therefore, we have modeled the orthorhombic crystal with equivalent $ab$ planes
separated by approximately 14.4 \AA, keeping the same geometry along the $ab$ plane. This
inter-planar distance preserves weak chemical bonds across neighbor planes and similar dielectric
screening as in the staggered stacking, thus ensuring equivalent electronic properties between the
model crystal and the crystal as characterized by X-ray measurements \cite{Jurchescu06}.  We
obtained a DFT-GGA band gap of 1.20~eV for the model unit cell and 1.14~eV for the experimental
structure~\cite{Jurchescu06}. We also calculate the diagonal components of the static dielectric
tensor $\epsilon_\infty$ within the random phase approximation (RPA), obtaining 2.55, 2.83 and 3.12
for $a$, $b$ and $c$ polarizations respectively. Spectroscopic determinations for these quantities
are: 2.6, 3.2 and 2.2~\cite{Tavazzi07_PRB} for $a$, $b$, and $c$ polarizations (2.15, 2.4 for $a$
and $b$~\cite{Li07}) respectively~\footnote{We follow
  the axes notation used in Ref. ~\cite{Li07}.}. The agreement with results from the experimental structure~\cite{Jurchescu06} demonstrates that the model unit cell is sufficient for evaluating the quasi-particle energies and the excitons. The small difference along $c$ is a consequence of the reduced unit cell. 

In the framework of the BSE method, we calculate the imaginary part of dielectric functions from
the optical transition matrix:
\begin{equation}
\epsilon_2(\omega) = \frac{4\pi^2 e^2}{\omega^2}\sum_S \left| \sum_{v,c}
 A^S_{vc}\langle v|{\vec \lambda}\cdot{{\vec V}}|c\rangle\right|^2\delta(\omega - \Omega_S)
\label{eps2}
\end{equation}
where $v$, $c$ are the single-particle valence and conduction states, ${\vec V}$ is the velocity
operator, ${\vec \lambda}$ is the polarization of light, $\Omega_S$ are the excitation energies of
the excited states $S$ in the crystal, and $A^S_{vc}$ are the expansion coefficients of the excited
states in electron-hole pair configuration $|S\rangle = \sum A^S_{vc} a_c^\dagger a_v |0\rangle$.
The expansion coefficients are obtained as eigenvectors of the effective Hamiltonian in the BSE.
Detailed formalism and technique are described in~\cite{Rohlfing00}. We apply a $4\times8\times4$
sampling of the Brillouin zone and solve the BSE equations for 8 valence and 8 conduction bands.
The real part of the dielectric function is a Kramers-Kronig continuation of Eq.~(\ref{eps2}).
Finally, the refractive index is obtained as $n = \sqrt{ \epsilon_1 + i \epsilon_2 }$.

We neglect the effect due to electron-phonon coupling in this work as the excitonic effect arising from purely electronic orbitals is our main concern.  The approximation is justified by a recent study of the effective masses of carriers in rubrene crystal field effect transistor which showed that the dominant excitations are quasiparticles rather than polarons~\cite{Li07}. It is also corroborated by the spectroscopy data which suggested that primary excitons in rubrene crystals are free excitons which do not involve lattice vibrations~\cite{Najafov06}.
\begin{figure}
\includegraphics[width=8cm]{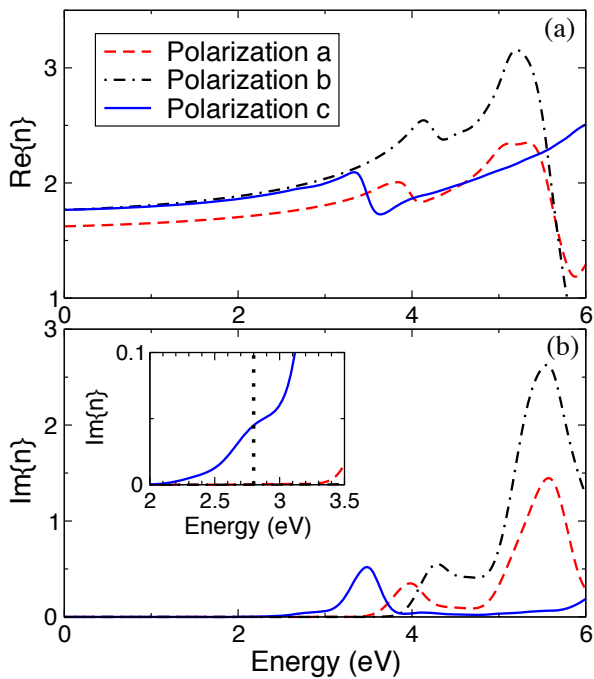}
\caption{Real (a) and imaginary (b) parts of the refractive index along the crystalline axes of the
rubrene crystal. The vertical line in the inset marks the GW quasiparticle gap.} 
\label{spectra}
\end{figure}

The imaginary part of the refractive index along all three crystalline directions of the rubrene
crystal is shown in Fig.~\ref{spectra}b. The first exciton line is at 2.3~eV, polarized along the
$c$ axis. Above it, there is a series of bound excitons with the same polarization, forming a flat
``hump'' at the onset of photo-absorption (see inset of Fig.~\ref{spectra}). This exciton band
should be associated to the first singlet level of the rubrene molecule coming from the HOMO-LUMO
molecular transition and it is in good agreement with the experimentally determined absorption onset~\cite{Tavazzi07_PRB,Mitrofanov06,Najafov06}.  The oscillating features in the experimental spectra correspond to vibrational resonance which is not included in the present theory. Tavazzi {\it et al.}~\cite{Tavazzi07_PRB} have reported weak absorption peaks around 2.5~eV in the $ab$ plane and attributed them to the second singlet of rubrene molecule in the $LN$ plane. This interpretation is not supported by our results. We find no remarkable oscillator strength around this energy with in-plane polarizations. Indeed, our results for the rubrene molecule show that the second singlet is two orders of magnitude weaker than the first one. The absorption lines reported at 2.5~eV in the $ab$ plane could also be enhanced by crystal imperfections or coupling to phonon bands, effects which are not included in our study. Our calculations further show absorption peaks at about 3.9 and 4.1~eV while the experimental peaks appear at about 3.7 and 4.0~eV along $a$ and $b$ axes, respectively.

The real part of the refractive index is shown in Fig.~\ref{spectra}a. A pronounced anisotropy can
be seen with enhanced response along the $b$ axis, in agreement with
experiment~\cite{Tavazzi07_PRB, Li07}. For the $a$ and $b$ polarizations, diagonal components of
the static dielectric tensor are found to be 2.62 and 3.1, larger than the values we obtained using
the random phase approximation (RPA) and closer to those obtained from
experiment~\cite{Tavazzi07_PRB}. Our GW-BSE value is larger than the measured one along
$c$~\cite{Tavazzi07_PRB}, for the same reason as in the RPA calculations discussed earlier.

The GW approximation predicts a quasi-particle band gap of about 2.8~eV,
increased from the DFT-GGA gap of 1.2~eV~\cite{Li07}. The difference
between this gap and the calculated excitation energies gives rise to a
binding energy of 0.5~eV for the first singlet exciton. The binding
energy of the first exciton is comparable to those reported for
pentacene (0.3-0.5~eV~\cite{Tiago03}) an tetracene
(0.4~eV~\cite{Hummer05}). No direct measurements of the electronic gap
of rubrene crystal could be found in the literature. Furthermore, there
is evidence that photo-excitation does not create uncorrelated
electron-hole pairs directly~\cite{Najafov06}, suggesting it is
difficult to directly measure the electronic gap without pulling
electrons and holes apart. This behavior is typical of tightly bound
excitons. Our study, combined with the experiments, thus allow us to provide additional information on the underlying band structure of rubrene. 

\begin{figure}
\includegraphics[width=8cm]{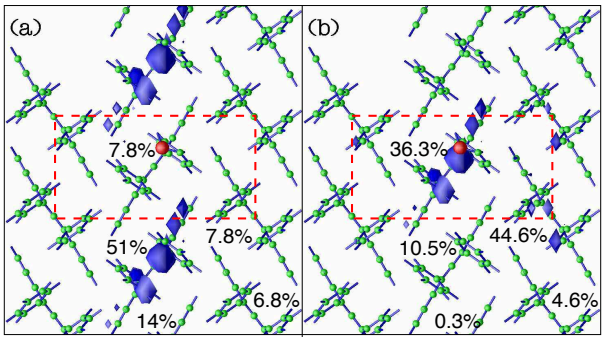}
\caption{Electron-hole probability distributions of the first spin-singlet (a) and spin-triplet (b) excitons of rubrene crystal. Red dot in the center of the unit cell (viewed along c axis) marks the hole while the isosurfaces indicate the distribution of the electron. The probabilities for finding the electron on each molecular site are indicated by the percentage.} 
\label{excitons}
\end{figure} 
Spin-triplet excitons are not accessible in optical absorption experiments, nevertheless they play
a fundamental role in emission processes such as phosphorescence~\cite{Pope}. The long life-time of triplet excitons is a consequence of the electron and hole having parallel spins. This also cancels out the exchange repulsion, raising the binding energy. The first spin-triplet exciton in organic molecular crystals is often a molecular exciton with very high binding energy~\cite{Davydov,Tiago03}. Rubrene is not an exception. The triplet has excitation energy 1.4~eV, close to the QMC excitation energy of the first triplet exciton in gas-phase rubrene. Fig.~\ref{excitons} shows that the first triplet exciton is a fairly well-defined molecular exciton, i.e., electron and hole have a 36\% probability of being found at the same molecular site. 

In contrast, the probability of electron and hole being at the same molecular site is just 8\% for the singlet (Fig.~\ref{excitons}a). The electron and hole are primarily distributed over second neighbor molecular sites, separated by approximately 7~\AA~\footnote{ Using $-e^2/\epsilon r$ with an average dielectric constant $\epsilon \sim 2.5$ and the interatomic distance $r = 7$ \AA, we obtain a binding energy of 0.8~eV, slightly larger than the GW/BSE value.} indicating it is a charge-transfer exciton. Excitons in the
2.3-2.9~eV range shown in Fig.~\ref{spectra} are also charge-transfer states but of larger size.
The exchange repulsion prevents the first spin singlet from being intramolecular. The exchange
repulsion is 2~eV when electron and hole are on the same site but drops to less than 0.1~eV when
electron and hole are on neighbor sites, owing to its short range. This drop compensates the drop
in Coulomb attraction between the on-site (2~eV) and the off-site configuration (1~eV). As a
result, an intramolecular spin-singlet exciton would be more energetic than the first
charge-transfer one by approximately 1~eV. This interpretation is supported by our gas-phase
calculations: assuming that the GW electron affinity is overestimated by 0.6 to 0.9~eV, as
indicated by QMC, we estimate the excitation energy of intramolecular spin-singlet exciton to be
$\sim3$~eV, i.e., 0.7~eV higher than the first charge-transfer exciton. 

To conclude, we have studied the electronic excitations
and optical spectra of crystalline and molecular rubrene from
first-principles approaches including both electron self-energy corrections
 and electron-hole correlations. We find the lowest dipole-allowed exciton at around 2.3~eV, both
for the molecule and the crystal, thus confirming the experimental absorption energies. These
excitons are spin-singlets and originate from transitions between the HOMO and LUMO. We also
achieved good agreement in the reflectance spectrum. The spatial distribution of the spin-singlet
and triplet excitons are analyzed and compared for the rubrene crystal. We have shown that the
optical spectrum and quasi-particle gap are strongly affected by many-body effects. Thus, the
results reported here are key to understand the electronic structure of the rubrene crystal.
Charge-transfer and molecular excitons might exist in other molecular crystals
\cite{Tiago03,Hummer05} and arrays of quantum dots. Moreover, charge-transfer excitons depend on
the local crystalline structure, therefore we predict a high sensitivity of the optical spectrum to pressure-induced phase transitions in this material.

Research supported by NSF grant No. DMR-0551195 (NS and JRC) and the Division of Materials Sciences and Engineering BES and the Solid State Lighting Program EERE, U.S. DOE under contract with
UT-Battelle, LLC (MLT and FAR). Computational support was provided by the Texas Advanced Computing
Center and the National Energy Research Scientific Computing Center. We would like to thank P.R.C
Kent for discussions.

\end{document}